\documentclass{article}

\usepackage{PRIMEarxiv}

\usepackage[utf8]{inputenc} 
\usepackage[T1]{fontenc}    
\usepackage{hyperref}       
\usepackage{url}            
\usepackage{booktabs}       
\usepackage{amsfonts}       
\usepackage{nicefrac}       
\usepackage{microtype}      
\usepackage{lipsum}
\usepackage{fancyhdr}       
\usepackage{graphicx}       
\graphicspath{{media/}}     
\usepackage{authblk}
\usepackage{graphicx,verbatim}
\usepackage{amsmath}    
\usepackage{bm}         
\usepackage{mathtools}
\usepackage{booktabs}   
\usepackage{graphicx}
\usepackage{hyperref}
\usepackage{algorithm}
\usepackage{multirow}
\usepackage{tabularx}
\usepackage{array}
\usepackage{graphicx,booktabs,multirow,makecell}
\usepackage{subfigure} 
\usepackage{array}
\usepackage{utfsym}
\usepackage{xcolor}
\usepackage{multirow}
\usepackage{graphicx,verbatim} 
\usepackage{amssymb}       
\usepackage{mathtools}
\usepackage{booktabs}   
\usepackage{graphicx}
\usepackage{hyperref}
\usepackage{multirow}
\usepackage{tabularx}
\usepackage{colortbl}
\usepackage{array}
\usepackage{ulem}
\usepackage{algpseudocode}
\usepackage{times}      
\usepackage{float}
\usepackage{caption}
\usepackage{textcomp}
\usepackage{amsmath}
\usepackage{multirow}
\usepackage{xcolor}
\usepackage{booktabs}
\usepackage{makecell}
\usepackage{array}
\usepackage{threeparttable}
\usepackage[authoryear]{natbib}
\title{STARFormer: A Novel Spatio-Temporal Aggregation Reorganization Transformer of FMRI for Brain Disorder Diagnosis}
\author[1,$\dagger$]{Wenhao Dong}
\author[1,3,$\dagger$]{Yueyang Li}
\author[1,*]{Weiming Zeng}
\author[1]{Lei Chen}
\author[2]{Hongjie Yan}
\author[3]{Wai Ting Siok}
\author[3,*]{Nizhuan Wang}
\affil[1]{Laboratory of Digital Image and Intelligent Computation, College of Information Engineering, Shanghai Maritime University, Shanghai 201306, China}
\affil[2]{Department of Neurology, Affiliated Lianyungang Hospital of Xuzhou Medical University, Lianyungang, China}
\affil[3]{Department of Chinese and Bilingual Studies, The Hong Kong Polytechnic University, Hong Kong SAR, China}
\affil[$\dagger$]{Equal contribution}
\affil[*]{Correspondence: zengwm86@163.com; wangnizhuan1120@gmail.com}

\begin{document}
\maketitle

\begin{abstract}
Many existing methods that use functional magnetic resonance imaging (fMRI) to classify brain disorders, such as autism spectrum disorder (ASD) and attention deficit hyperactivity disorder (ADHD), often overlook the integration of spatial and temporal dependencies of the blood oxygen level-dependent (BOLD) signals, which may lead to inaccurate or imprecise classification results. To solve this problem, we propose a Spatio-Temporal Aggregation Reorganization Transformer (STARFormer) that effectively captures both spatial and temporal features of BOLD signals by incorporating three key modules. The region of interest (ROI) spatial structure analysis module uses eigenvector centrality (EC) to reorganize brain regions based on effective connectivity, highlighting critical spatial relationships relevant to the brain disorder. The temporal feature reorganization module systematically segments the time series into equal-dimensional window tokens and captures multiscale features through variable window and cross-window attention. The spatio-temporal feature fusion module employs a parallel transformer architecture with dedicated temporal and spatial branches to extract integrated features. The proposed STARFormer has been rigorously evaluated on two publicly available datasets for the classification of ASD and ADHD. The experimental results confirm that STARFormer achieves state-of-the-art performance across multiple evaluation metrics, providing a more accurate and reliable tool for the diagnosis of brain disorders and biomedical research. The official implementation codes are available at: \url{https://github.com/NZWANG/STARFormer}.
\end{abstract}
\section{Introduction}

The human brain function can be characterized by intricate functional networks where multiple brain regions cooperate to facilitate cognitive processes and mental states \citep{zhang2024learning}. Disruptions of these networks can manifest as various neurodevelopmental conditions, such as autism spectrum disorder (ASD) \citep{bicks2024functional} and attention deficit hyperactivity disorder (ADHD) \citep{koirala2024neurobiology}. However, the underlying mechanisms of these disruptions are not yet fully understood \citep{pievani2011functional,wee2014disrupted,qian2019large}. Resting-state functional magnetic resonance imaging (rs-fMRI) has emerged as a powerful, noninvasive technique for investigating these functional networks by measuring blood oxygen-level-dependent (BOLD) signals with relatively high spatial and temporal resolution \citep{Pei2025bold}. In particular, functional connectivity (FC), which reflects the temporal correlations of BOLD signals between regions of interest (ROI), provides crucial insights into brain organization and cognitive functions \citep{li2024mhnet,Zhou2024fusing}.

Traditional diagnostic approaches, which primarily rely on symptomatic observations and clinical expertise, are inadequate for detecting intricate patterns across the entire brain. This limitation highlights the need for more advanced analytical methods. Initially, traditional machine learning techniques were used to analyze multivariate brain responses from rs-fMRI data for diagnosing neurodevelopmental disorders \citep{khosla2019machine}. The field then progressed with the advent of deep learning \citep{shoeibi2023diagnosis} and, more recently, the groundbreaking introduction of transformer models \citep{vaswani2017attention}. Transformers have significantly improved the modeling of complex patterns in high-dimensional data through self-attention mechanisms, offering better scalability and more efficient capture of global network information than conventional approaches \citep{cong2024comprehensive,Mubonanyikuzo2024DetectionOA}. However, traditional transformer models often struggle to capture both the intricate spatial and temporal features within fMRI data simultaneously. The standard self-attention mechanism in transformer encoders tends to focus on identifying parts of the time series with similar patterns across the entire sequence, especially those with matching peaks. This approach may overlook patterns that are similar in the short to medium term, as they occur closely in time. In fMRI data analysis, recognizing these short- to medium-term patterns is crucial because significant changes in BOLD signals may only occur within shorter time windows, rather than being consistent throughout the entire duration \citep{hutchison2013dynamic}.

To address these limitations, we propose a novel spatio-temporal aggregation reorganization transformer (STARFormer) that effectively captures both spatial and temporal information of brain functional networks through a designed dual-branch architecture. This approach uses effective connectivity to provide spatial information for the time series of BOLD signals. Unlike methods that rely solely on FC, this approach preserves temporal dynamics while incorporating spatio-temporal information. STARFormer, through ROI rearrangement and a variable window strategy, not only improves diagnostic accuracy but also reduces computational complexity by leveraging windowed computations. Compared with convolution-attention hybrid architecture of LCGNet \citep{zhou2024lcgnet}, dual-branch design of STARFormer separates and integrates spatiotemporal features more effectively, while providing stronger interpretability via ROI analysis. These improvements allow STARFormer to achieve more accurate diagnosis and better clinical interpretability while maintaining computational efficiency.

For spatial information modeling, we specifically choose eigenvector centrality (EC) over other centrality measures based on several critical considerations \citep{EC}. First, EC measures the global influence of a node within the entire network-it depends not only on the node's direct connections but also on the centrality of the nodes it connects to. Second, the Transformer architecture inherently lacks spatial structure inductive bias and requires externally provided ordering or positional information to learn the organizational patterns between nodes. EC not only identifies globally important nodes in brain networks but also provides a stable and coherent spatial ordering method. This biologically grounded approach effectively incorporates the network roles of ROIs into the input structure, guiding the Transformer to focus on key brain regions and thereby enhancing brain disorder classification performance.

Our methodology introduces three key innovations: 

1) \textbf{ROI Spatial Structure Analysis Module}: This module employs EC to reorganize brain regions based on their functional importance within seven established brain networks, ensuring the preservation and enhancement of crucial spatial relationships. This reorganization significantly improves the model's ability to identify disorder-related spatial patterns. 

2) \textbf{Temporal Feature Reorganization Module}: This module integrates multiscale local features with global representations through a unique variable window strategy, enabling the model to capture fine-grained temporal patterns at different scales while maintaining computational efficiency. The implementation of cross-window attention enhances the model's capacity to capture both short-term and long-term temporal dependencies in time series.

3) \textbf{The Spatio-Temporal Feature Fusion Module}: This module comprises temporal and spatial branches that simultaneously extract multiscale temporal dependencies and spatial representations. The temporal branch incorporates the temporal feature reorganization module to learn local and global temporal features, while the spatial branch uses the reorganized ROI structure to capture disorder-specific patterns. 

These innovations collectively make STARFormer a powerful tool that significantly enhances the accuracy and efficiency of diagnosing brain disorders by capturing critical spatio-temporal features. Overall, the contributions of our work are summarized as follows.
\begin{itemize}
	\item A transformer architecture is proposed to enhance brain-disorder diagnosis by integrating disorder-specific ROI spatial information, thereby improving the precision and efficiency of fMRI analysis.
	\item A novel variable window-based temporal feature reorganization module is included, allowing STARFormer to capture both local and global features by adjusting window size.
	\item A spatio-temporal feature fusion module is developed to fully explore deep spatio-temporal features, enabling comprehensive feature representation.
\end{itemize}

\section{Related Work}
\subsection{Centrality Nodes Identification in Brain Disorders}
Identifying critical brain nodes is crucial for understanding disorder-specific functional disruptions, as neurological disorders significantly differ in regional connectivity patterns \citep{van2019cross}. For example, individuals with mild cognitive impairment exhibit altered effective connectivity in memory-related areas such as the hippocampus and amygdala \citep{zheng2017altered}, while individuals with ASD display distinctive alterations in regions associated with social cognition, such as the default mode network \citep{monk2009abnormalities}.

Measures of node centrality have been effectively applied in diagnosing brain disorders to uncover differential patterns. For example, Saha \citep{saha2024eigenvector} used EC to assess group differences in the centrality of brain regions, distinguishing between typically developing children and those with ASD. Similarly, Grobelny et al. \citep{grobelny2018betweenness} utilized betweenness centrality (BC) to construct diagnostic models for pediatric epilepsy, revealing that nodes with high BC during seizure onset could represent centers in self-regulating networks that help terminate seizures. Liao et al. \citep{liao2021changes} found that changes in degree centrality (DC) in Parkinson's disease were frequency-related and frequency-specific. 

However, due to computational complexity, BC, DC and other centrality measurement methods are less suitable for analyzing full-brain networks that involve a large number of voxels \citep{saha2024eigenvector}. Hence, this study focuses on EC to identify differential connectivity patterns, highlighting the importance of key nodes in distinguishing pathological alterations.

\subsection{Features of fMRI for Brain Disorders Identification}
Besides directly exploring the relationship between node centrality and brain disorders, the extraction of FC features is also a key focus in many brain disorder identification studies. A typical approach for identifying brain disorders involves extracting FC, which represents the temporal correlation matrix of BOLD signals from ROI, followed by the use of classifiers such as SVM and logistic regression to reduce dimensionality \citep{khosla2019machine}. A study by \citep{zhou2024lcgnet} proposed a novel architecture called local sequential feature coupling global representation learning (LCGNet), which uses convolution operations and self-attention mechanisms to enhance representation learning in fully convolutional networks for automatic brain disorder classification. 

However, FC is limited to capturing linear relationships and lacks causal or directional insights. The effective connectivity approach has emerged as a promising tool that describes brain activity by incorporating causal interactions between brain regions \citep{wang2024deep}. The study by Dai et al. \citep{dai2023altered} indicated that patients with major depressive disorder (MMD) exhibit significantly altered effective connectivity networks in various brain networks during the resting state, which may serve as potential biomarkers. Unlike FC, which focuses solely on statistical correlations, effective connectivity elucidates the directional influence between regions, providing a deeper understanding of the connections within the brain \citep{sharaev2016effective}.

\begin{figure*}
	\centering
	\includegraphics[width = 1\textwidth]{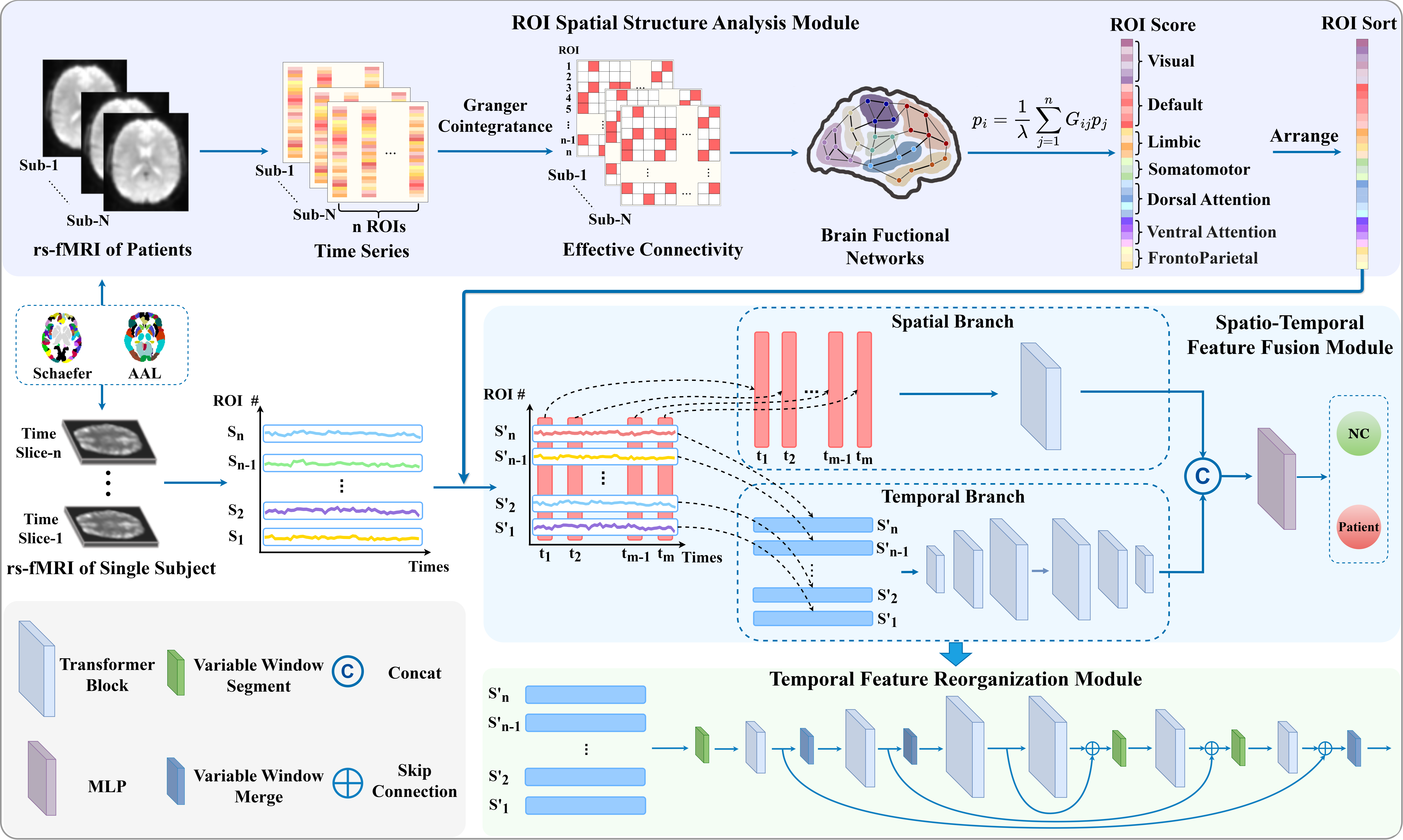}
	\caption{Architecture of STARFormer in fMRI data for brain disorder diagnosis.}
	\label{STARFormer}
\end{figure*}

Although deep learning methods based on FC have been widely used for diagnosing brain diorders, some studies reveal that these methods can overlook the temporal nature of fMRI data, thereby losing vital information on temporal dynamics \citep{deng2022sttransformer}. Recent advances in time-domain-based deep learning methods have shown great promise in diagnosing brain disorders. For instance, long short-term memory (LSTM) networks have been used to classify individuals with ASD and typical controls from multi-site fMRI time series \citep{dvornek2017lstm}, while a transformer model with a fusion window has further advanced fMRI time series analysis \citep{bedel2023bolt}.

\subsection{Spatio-Temporal Information Integration for Brain Disorders Identification}
For modeling fMRI features, most deep learning models use features like FC but do not fully exploit the spatial and temporal dependencies of fMRI signals, which limits the precise analysis of brain activity. The gradient-weighted Markov random field (gwMRF) model mentioned in \citep{wang2024multi} enables spatially-ordered brain region extraction. This study demonstrated that spatial disorganization decreases model performance, emphasizing the need to preserve spatial information in rs-fMRI. Recent studies increasingly emphasize the importance of spatial-temporal information integration. For example, the use of 3D-CNNs can extract spatial features from fMRI data for diagnosing ASD and ADHD \citep{li20182, zou20173d}. Liu et al. \citep{liu2023stcal} proposed STCAL, a spatio-temporal cooperative attention learning model that employs a guided cooperative attention module to simultaneously capture spatio-temporal correlations and learn fine-grained attention representations from time series fMRI data. Zhang et al. \citep{tomography10120138} used independent component analysis (ICA) to aggregate spatio-temporal information from fMRI, improving the performance of depression diagnosis.

Transformer models have also made significant advances in integrating spatio-temporal features, showcasing their potential to capture complex dependencies within fMRI data. In \citep{deng2022sttransformer}, a novel transformer-based framework, the ST-Transformer, was introduced, featuring a linear spatio-temporal multihead attention unit to extract spatio-temporal features from fMRI data through spatial self-attention. Similarly, a spatio-temporal graph transformer network was proposed in \citep{he2024stgtn}, incorporating a spatial transformer-based graph message-passing mechanism to capture inter-regional relationships and using FC as edge features.

Although existing studies have advanced brain disorder diagnosis, they exhibit critical limitations. Many methods ignore temporal dynamics of fMRI signals or overlook spatial structure relationships, while few provide biologically interpretable explanations for clinical adoption. To address these limitations, STARFormer directly utilizes fMRI time-series data to capture temporal dynamics, integrates spatial relationships through EC-based spatial feature reorganization to enhance both spatial structure learning and biological interpretability for more accurate and clinically relevant diagnosis.

\section{Method}
An overview of STARFormer combining the spatial and temporal information is illustrated in Fig.\ref{STARFormer}. It consists of the ROI spatial structure analysis module, the temporal feature reorganization module and the spatio-temporal feature fusion module.

\subsection{ROI Spatial Structure Analysis Module}
The spatial information of the brain is provided by the effective connectivity matrix $\mathbf{G}$ of N patients. We use Granger causality (GC) to calculate effective connectivity between different brain regions, as it can reveal whether the time series of a brain region  predict the time series of another \citep{shojaie2022granger}. Suppose the chosen atlas divides the brain into $n$ ROIs, each with a time series $s_i(t)$ of length $m$. We first establish an autoregressive model, where the time series of each ROI $i$ can be predicted using its own values from the past $h$ time points:
\begin{equation}
	s_i(t)=\sum_{k=1}^ha_{ik}s_i(t-k)+\varepsilon_i(t),
\end{equation}
where $a_{ik}$ is the regression coefficient and $\varepsilon_i(t)$ is the prediction error or residual. 

Then, a bivariate GC model is constructed to test for Granger causal effects between ROI:
\begin{equation}
	s_j(t)=\sum_{k=1}^ha_{jk}s_j(t-k)+\sum_{k=1}^hb_{ik}s_i(t-k)+\varepsilon_j'(t),
\end{equation}
where $b_{ik}$ is the coefficient of influence of node $i$ on node $j$. Subsequently, the residual variances of both models are computed and an F-test is used to evaluate whether the models exhibit a statistically significant difference. A significant F-value indicates that the time series of ROI $i$ Granger-causes the time series of ROI $j$. For each pair of ROI $(i, j)$, the results of the GC test are indicated as $G_{ij}$, where $G_{ij} = 1$ indicates a GC relationship from ROI $i$ to ROI $j$, while $G_{ij} = 0$ indicates the absence of such a relationship. Finally, a valid connection matrix $\mathbf{G}$ for $n \times n$ can be constructed as
\begin{equation}
	\mathbf{G}=\begin{pmatrix}0 & G_{12} & \cdots & G_{1n}\\ G_{21} & 0 & \cdots & G_{2n}\\ \vdots & \vdots & \ddots & \vdots\\ G_{n1} & G_{n2} & \cdots & 0\end{pmatrix}.
\end{equation}

The ROIs and the connectivity metrics of the effective connectivity matrix $\mathbf{G}$ define the nodes and edges of the graph, respectively. Serving as the adjacency matrix of the network, $\mathbf{G}$ facilitates the assessment of node centrality through its eigenvectors.

We chose EC to identify key brain regions because it more accurately captures global influence while remaining computationally efficient and theoretically aligned. Unlike DC, which merely counts direct connections, EC recursively weights each link by the importance of its neighbors to reflect a node's true influence across the entire network. In contrast to BC-whose O(n³) complexity is impractical for full‐brain networks of hundreds of ROIs—and closeness centrality(CC)-which has theoretical limitations in directed networks—EC remains both efficient and robust. Crucially, its iterative calculation naturally models the propagation of causal influences, making it ideal for pinpointing hub regions whose altered connectivity underlies disease.

The EC score $p_i$ for each ROI $i$ measures its connection strength to other influential ROIs in the network and can be expressed as
\begin{equation}
	p_{i}=\frac1\lambda\sum_{j=1}^{n}G_{ij}p_{j},
\end{equation}
where $\lambda$ is the largest eigenvalue of $\mathbf{G}$, and $\mathbf{G}_{ij}$ represents the connectivity measure from ROI $i$ to ROI $j$.

The EC problem can be expressed in matrix form as
\begin{equation}
	\mathbf{GP} = \lambda \mathbf{P},
\end{equation}
where $\mathbf{P} = [p_1,p_2,...,p_n]^T \in R^{n \times 1}$ represents the EC vector, with $\lambda$ denoting its corresponding eigenvalue. The calculation requires determining the principal eigenvector $\mathbf{P}$ of the adjacency matrix $\mathbf{G}$ in conjunction with its dominant eigenvalue $\lambda$. Subsequently, the EC of any given ROI $i$ is precisely captured by the component $p_i$ within this principal eigenvector associated with the maximal eigenvalue $\lambda$. To guarantee uniqueness and stability, $p_i$ is normalized such that
\begin{equation}
	p_{i}=\frac{p_{i}}{\sum_{i=1}^{n}p_{i}},
\end{equation}
thereby normalizing the centrality values.

Following the computation of the EC vector for ROIs of each patient, the average EC vector $\mathbf{\bar{P}}\in R^{n \times 1}$ for all patients was determined by
\begin{equation}
    \begin{gathered}
        \bar{p_{i}}=\frac1N{\sum_{i=1}^{N}p_{i}},\\
	\mathbf{\bar{P}} = [\bar{p_1},\bar{p_2},...,\bar{p_n}]^T.
    \end{gathered}
\end{equation}

We consider using the obtained EC score $\bar{p_i}$ of ROI to rearrange the spatial information. To reduce noise and errors caused by unrelated regional arrangements, ROIs are grouped according to the seven functional networks of the brain (visual network, somatomotor network, default network, limbic network, dorsal attention network, ventral attention and frontoparietal network) \citep{yeo2011organization}. 
A hierarchical ranking strategy not only prevents functionally related regions from being scattered but also captures disease‐specific important ROIs, producing an ROI sequence that is more robust to disease variability and noise.
The ROIs within each group are then arranged by $\bar{p_i}$ in descending order:
\begin{equation}
	\begin{gathered}
		Network_i = \{ROI_1, ROI_2, \dots, ROI_r\}, \\
		Network_{i}^{'} = Arranging(Network_{i}),
	\end{gathered}
\end{equation}
where $Network_i$ is the i-th functional network, $r$ denotes the number of ROIs within the functional network. The final sorting result $Sort$ is expressed as follows:
\begin{equation}
	Sort=\{Network_{1}^{'},Network_{2}^{'},...,Network_{7}^{'}\}.
\end{equation}

The global EC of ROIs allows for better capture of interactions or dependencies between regions. Rearranging ROIs aligns the order of the brain network with the spatial characteristics of the patients, improving the stability and consistency of its analysis. The sorting results of the ROIs are applied to the fMRI time series data for all participants $S=\{S_1(t),S_2(t),...,S_n(t)\} \in R^{n \times m}$, and
\begin{equation}
	S' = Ordering(S,Sort),
\end{equation}
where $S'= \{S'_1(t),S'_2(t),...,S'_n(t)\}$ represent the fMRI time series data after the application of ordering.

\begin{figure*}
	\centering
	\includegraphics[width = 1\textwidth]{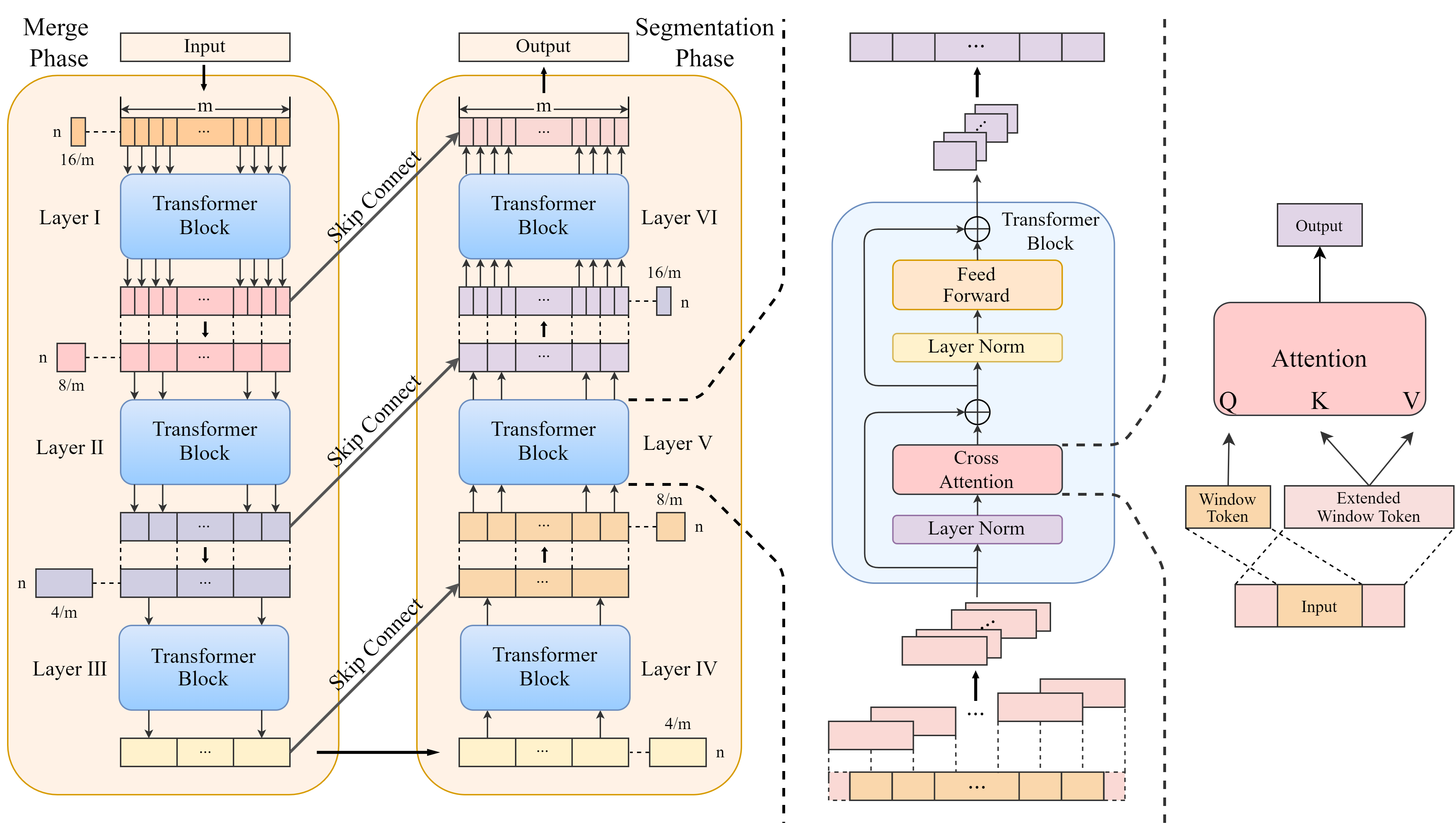}
	\caption{Architecture of the temporal feature reorganization module in STARFormer, which employs variable window and cross-window attention to capture both local temporal patterns and global dependencies in fMRI data.}
	\label{win}
\end{figure*}

\subsection{Spatio-Temporal Feature Fusion Module}

The spatio-temporal feature fusion module employs a parallel network architecture consisting of temporal and spatial branches, which extract multi-level features from fMRI data without mutual interference. The temporal branch operates through a dual-phase mechanism of merge and segmentation, where input data undergoes segmentation into window tokens processed through variable windows, facilitating inter-window information exchange via cross-window attention mechanisms. Meanwhile, the spatial branch maintains a fundamental transformer architecture with standard self-attention mechanisms for spatial information extraction. This enables STARFormer to effectively leverage the node spatial information derived from the ROI spatial structure analysis module to uncover disorder-related latent spatial representations.

Arrayed rs-fMRI data $S'$ are used as input to the spatio-temporal feature fusion module. $S'$ enter the temporal branch and the spatial branch, respectively. The temporal branch treats each row $S'_i(t)$ in $S'$ as a token, allowing the cross-window attention in the temporal branch to learn local features. When inputting into the spatial branch, the spatial branch adjusts the dimensions of $S'$ to $m \times n$ as the new input $S^T=\{t_{1},t_{2},...,t_{m}\}\in R^{m \times n}$, treating each column $t_i$ in $S'$ as a token. Self-attention in the spatial branch can learn the global dependencies between brain regions. The results obtained from the two branches are then integrated, and the detailed operations can be expressed as
\begin{equation}
	F=Concat[Sp(S'),Te(S^T)^T],
\end{equation}
where $Sp$ refers to the processing of the spatial branch and $Te$ refers to the processing of the temporal branch. Finally, the combined feature $F$ was input into the multilayer perceptron (MLP) for classification.
The MLP has two layers, uses ReLU activation, and a hidden dimension of 256.

\subsection{Temporal Feature Reorganization Module}

Exclusive reliance on global similarity metrics is hypothesized to obscure critical time-dependent local patterns, potentially diminishing ability of the model to capture intricate features within fMRI data. To enhance classification accuracy while reducing computational complexity, this work introduces a novel Temporal Feature Reorganization Module based on variable window.

\subsubsection{Window Tokens and Extended Window Tokens}
In the temporal branch, the arranged fMRI time series data $S'$ are segmented into equally sized window tokens $\{x_{1}^{l},x_{2}^{l},...,x_{g}^{l}\}$ along the temporal dimension by the variable window, where $g$ is the number of window tokens and $l$ is the layer of the temporal branch. The length of each window token $x$ is $w=m/g$. The extended window token $y_i$ is formed by extending $x_i$ at both the start and the end of the temporal dimension by a length of $w/2$, respectively. The length of $y_i$ is $2w$. To ensure that the lengths of the extended window tokens are consistent, padding of length $w/2$ is added to the start of the first window token $x_1$ and the end of the last window token $x_g$ to form the extended window tokens $y_1$ and $y_g$, while the overall length of the time series extends to $m+w$. It is intended that the positions of this padding do not participate in the subsequent backpropagation process, so zero vectors of length w/2 will be used to cover the padded positions during training to avoid affecting the performance of the model.

\subsubsection{Transformer Block}
The window tokens will serve as inputs to the transformer block. The transformer block comprises a layer norm (LN) layer, a cross attention layer, and a feed forward (FF) layer. The LN layer is responsible for applying layer normalization to the input data, which improves the model training process. To facilitate cross-window information interaction between window tokens, the model employs cross-window attention between window tokens instead of self-attention within window tokens. The cross attention layer receives extended window tokens of length $2w$ as input. When processing a time series composed of $g$ window tokens, let $Q$ represent the queries in the attention mechanism, and $K$ and $V$ represent the keys and values, respectively. The query, key, and value can be expressed as
\begin{equation}
	\begin{split}
		\mathrm{Q}&=f_q(x_1,x_2,...,x_g), \\
		\mathrm{K}&=f_k(y_1,y_2,...,y_g), \\
		\mathrm{V}&=f_v(y_1,y_2,...,y_g),
	\end{split}
\end{equation}
where $f_q,f_k,f_v$ are learnable linear projections.

Performing global self-attention requires calculating the relationships between each token and all other tokens, which undoubtedly involves high computational complexity. However, using the extended window in the attention computation allows the representation of local information to no longer be limited to its own window tokens. This facilitates the interaction of information between window tokens and reduces computational complexity. To better capture specific positional information between window tokens, bias is incorporated into the attention computation to adjust the attention weight matrix \citep{liu2021swin, yang2021focal}:
\begin{equation}
\label{eq:attention}
	Attention(Q_{},K_{},V_{})=Softmax(\frac{Q_{}K_{}^{T}}{\sqrt{d}}+\mathbf{B})V,
\end{equation}
where $\mathbf{B}$ is a learnable positional bias matrix and $d$ is the feature dimension of the attention heads. $\mathbf{B}$ represents the positions of the window tokens relative to all the tokens in the receptive field, including both the window tokens and the extended window tokens. The FF layer applies a nonlinear transformation to the input and utilizes Dropout to prevent model overfitting. The activation function used is GELU (Gaussian error linear unit) \citep{hendrycks2016gaussian}.

\subsubsection{Variable Window}
In the merge phase of the temporal branch, $S'$ will first be segmented into 16 equal-length window tokens $\{x_{1}^{1}, x_{2}^{1}, ..., x_{16}^{1}\}$ and input into the transformer block for computation. In the next layer, the computed results will be sequentially merged in pairs into 8 equal-length window tokens $\{x_{1}^{2}, x_{2}^{2}, ..., x_{8}^{2}\}$, which will then be input into the transformer block for further computation. Afterward, the resulting outputs will be sequentially merged into 4 equal-length window tokens $\{x_{1}^{3}, x_{2}^{3}, x_{3}^{3}, x_{4}^{3}\}$ for computation in the subsequent transformer block. The merge phase is then expressed as

\begin{equation}
	\begin{gathered}
		x_{1}^{'l},x_{2}^{'l},...,x_{g}^{'l}=TransformerBlock(x_{1}^{l},x_{2}^{l},...,x_{g}^{l}), \\
		x_{i}^{l+1}=(x_{2i-1}^{'l},x_{2i}^{'l}),i=1,2,...,g/2.
	\end{gathered}
\end{equation}

In the merge phase, the number of layers $l\in\{1,2,3\}$ and the number of window tokens $g\in\{16,8,4\}$.

After entering the segment phase, the transformer block will first take the results from the previous layer as $\{x_{1}^{4}, x_{2}^{4}, x_{3}^{4}, x_{4}^{4}\}$ and perform computations again. In the next layer, the computed results will be segmented into 8 window tokens $\{x_{1}^{5}, x_{2}^{5}, ..., x_{8}^{5}\}$ for further computation. Finally, the 8 window tokens will be segmented into 16 window tokens $\{x_{1}^{6}, x_{2}^{6}, ..., x_{16}^{6}\}$ for the final computation. Additionally, skip connections are used between window tokens of the same size in both the merge and segment phases to alleviate the issues of gradient vanishing and explosion, accelerate model convergence, and reduce the complexity of the network. The segment phase is then expressed as
\begin{equation}
	\begin{gathered}
		x_{1}^{'l},x_{2}^{'l},...,x_{g}^{'l}=TransformerBlock(x_{1}^{l},x_{2}^{l},...,x_{g}^{l}),\\
		(x_{2i-1}^{l+1},x_{2i}^{l+1})=x_{i}^{'l}+x_{i}^{'7-l},i=1,2,...,g.
	\end{gathered}
\end{equation}

In the segment phase, the number of layers $l\in\{4,5,6\}$ and the number of window tokens $g\in\{4,8,16\}$. 

To quantify the computational benefits of our cross-window attention mechanism, we provide a rigorous complexity comparison with global self-attention. For a sequence of length $m$ with hidden dimension $n$, global self-attention requires computing attention weights between all token pairs, yielding time complexity $\mathcal{O}(m^2 n)$. In contrast, our cross-window attention divides the sequence into $g$ windows of size $w = m/g$, where each window processes extended tokens of size $2w$ to capture cross-window dependencies. This results in time complexity $\mathcal{O}(g \times (2w)^2 n) = \mathcal{O}(4m^2 n/g)$, achieving a computational reduction factor of $g/4$ compared to global self-attention.

The following Algorithm \ref{alg1} shows the procedure of spatio-temporal feature fusion module.

\begin{algorithm}
	\caption{Pseudocode of the spatio-temporal feature fusion module}
	\begin{algorithmic}
		\Require The time series data of the i-th subject $S'\in R^{n \times m}$
		\Ensure The predicted probability of the i-th subject set Prob
		{\textit{\# Spatial branch}}
		\State $\{t_1, t_2, ..., t_m\} \leftarrow (S')^T=\{S(t)_1', S(t)_2', ..., S(t)_n'\}^T$
		\State $\{t_1', t_2', ..., t_m'\} \leftarrow TransformerBlock\{t_1, t_2, ..., t_m\}$
		\State ${S^T}' = \{t_1', t_2', ..., t_m'\}^T$\newline
		{\textit{\# Temperal branch}}
		\State $S^1 \leftarrow S'$
		\For{$l = 1, 2, ..., 6$}  \textit{\#} {$l:$\textit{\ l-th layer of temperal branch}}
		\State $X^l = \{x_1^l, x_2^l, ..., x_g^l\} \leftarrow S^l$
		\State $Y^l = \{y_1^l, y_2^l, ..., y_g^l\} \leftarrow \{x_1^l, x_2^l, ..., x_g^l\}$
		\State $Compute(Q, K, V) \leftarrow LN(X^l), LN(Y^l), LN(Y^l)$
		\State $Attention\ Output \leftarrow CrossAttention(Q, K, V)$
		\State $Residual\ Connections \leftarrow X^l + Attention\ Output$
		\State $Norm \leftarrow LN(Residual\ Connections)$
		\State $S^l \leftarrow FF(Norm) + Residual\ Connections$
		\If{$l > 3$}
		\State $S^l \leftarrow S^l + S^{7-l}$
		\EndIf
		\EndFor
		\State $F = Concat(S^l, {S^T}')$
		\State $Output\ logits \leftarrow LinearTransformation(F)$
		\State $Prob \leftarrow Softmax(Output\ logits)$
	\end{algorithmic}
	\label{alg1}
\end{algorithm}

\section{Experiments}
\subsection{Datasets}
In this study, we conducted experiments using two public fMRI datasets: ABIDE-I and ADHD-200. The ABIDE-I dataset was compiled by collaboration between 17 international imaging sites, openly sharing 871 valid fMRI samples from 403 individuals with ASD and 468 typically developing controls \citep{craddock2013neuro}. The ADHD-200 dataset was collected from 8 international imaging sites, openly sharing 947 valid fMRI samples from 362 children and adolescents with ADHD and 585 typically developing controls \citep{bellec2017neuro}.

The preprocessed fMRI datasets are available in C-PAC of ABIDE-I and Athena of ADHD-200 \citep{cpac,bellec2017neuro}. Specifically, the preprocessing steps include voxel intensity normalization, motion correction, and slice timing correction. The fMRI images are then co-registered to their corresponding anatomical images and normalized to MNI152 space. Finally, the mean time series is extracted from each ROI for each subject on the basis of the specified atlas.  

\subsection{Experimental Process and Details}
We randomly selected 10\% of patient samples from the dataset as input into the ROI spatial structure analysis module. For each subject in the dataset, the fMRI time series was randomly cropped to 128 samples along the temporal dimension to maximize the retention of the sample information.
We chose the temporal length by considering the average duration of our samples to preserve as much information as possible. In addition, since each sample must be divided into an integer number of window tokens, we ultimately set the temporal dimension to 128.
ROI parcellation was determined using two public brain atlases: the Schaefer atlas \citep{schaefer2018local} and the AAL atlas \citep{tzourio2002automated}. 
The Schaefer atlas parcels the brain based on functional connectivity, emphasizing functional coherence, while the AAL atlas is organized by anatomical structure, facilitating comparison with traditional neuroanatomical studies. Using both atlases allows us to test our model's robustness under different spatial parcellation schemes. These two atlases are widely used in neuroimaging, making it easier to compare our results with existing work and to enhance interpretability and generalizability.
For the Schaefer atlas, we selected the scale of 400 ROIs across seven intrinsic connectivity networks. The AAL atlas partitions the brain anatomically into 116 ROIs. The STARFormer model was trained for 100 epochs with a batch size of 128 and a dropout rate of 0.5. Eight attention heads were specified, each with 16 dimensions. For training on ABIDE-I dataset, the initial learning rate was set to 5e-5, with a maximum of 1e-4, and gradually reduced to 1e-5. For training on ADHD-200 dataset, the initial learning rate was set to 1e-5, with a maximum of 5e-5, and finally reduced to 1e-6.

\begin{table*}[!ht]
	\centering
	\renewcommand{\arraystretch}{1.8}
        \footnotesize
	\caption{Performance comparison using the Schaefer atlas on ABIDE-I and ADHD-200 datasets.}
	\begin{tabular}{
			m{5.2cm}<{\centering}
			m{1.15cm}<{\centering}
			m{1.15cm}<{\centering}
			m{1.15cm}<{\centering}
			m{1.15cm}<{\centering}
			m{1.15cm}<{\centering}
			m{1.15cm}<{\centering}
			m{1.15cm}<{\centering}
			m{1.3cm}<{\centering}
		}
		\Xhline{1.2pt}
		& \multicolumn{4}{c}{\textbf{ABIDE-I}} & \multicolumn{4}{c}{\textbf{ADHD-200}} \\[1pt]
		\cline{2-9}
		\multirow{-2}{*}{\textbf{Model}} & \textbf{Acc} & \textbf{Prec} & \textbf{Rec} & \textbf{AUC} & \textbf{Acc} & \textbf{Prec} & \textbf{Rec} & \textbf{AUC} \\[1pt]
		\hline
		SVM \citep{abraham2017deriving} & 65.72±4.11 & 53.97±7.33 & 66.26±6.68 & 71.33±4.62 & 59.76±2.68 & 59.34±5.32 & 59.61±5.43 & 60.62±4.66 \\
		LSTM \citep{dvornek2017lstm} & 66.78±4.25 & 64.05±9.93 & 65.81±7.62 & 70.25±3.85 & 65.06±3.96 & 62.92±4.30 & 60.48±6.38 & 65.56±5.57 \\
		BrainNetCNN \citep{kawahara2017brainnetcnn} & 67.34±4.22 & 63.82±8.73 & 65.50±4.63 & 73.15±5.62 & 66.78±3.90 & 62.62±4.66 & 62.82±8.33 & 62.22±4.13 \\
		\hline
		SwinT \citep{liu2021swin} & 69.79±4.59 & 60.75±7.60 & 68.27±6.56 & 74.13±4.15 & 68.56±4.74 & 65.25±4.36 & 66.40±9.70 & 73.85±4.59 \\
		BolT \citep{bedel2023bolt} & 71.28±4.62 & 69.85±4.94 & 71.32±4.35 & 77.46±3.44 & 70.82±3.57 & 68.04±4.18 & 71.51±3.27 & 73.36±3.86 \\
		Com-BrainTF \citep{bannadabhavi2023community} & 72.75±4.56 & 70.65±4.54 & 78.43±4.63 & 77.93±3.02 & 68.94±2.68 & 67.36±4.23 & 72.92±3.88 & 73.54±3.68 \\
		\hline
		MAHGCN \citep{liu2023mahgcn} & 73.12±3.63 & 71.05±5.38 & 72.02±4.14 & 72.07±3.03 & 70.76±4.63 & 69.95±3.38 & 71.08±2.14 & 74.25±3.41 \\
		RGTNet \citep{wang2024rgtnet} & 74.43±4.82 & 73.67±4.38 & 75.28±4.06 & 74.55±2.73 & 72.03±3.11 & 70.14±3.81 & 71.33±4.45 & 74.28±2.82 \\
		PLSNet \citep{wang2023plsnet} & 75.17±4.62 & 75.91±3.65 & 79.82±5.83 & 77.03±2.17 & 72.53±4.27 & \textbf{73.69±3.82} & 72.42±3.22 & 78.02±2.87 \\
		\hline
		STARFormer & \textbf{77.57±3.70} & \textbf{76.98±3.27} & \textbf{84.38±3.50} & \textbf{78.29±3.23} & \textbf{74.12±2.47} & 73.37±3.07 & \textbf{73.54±3.27} & \textbf{78.81±2.18} \\
		\Xhline{1.2pt}
	\end{tabular}
	\label{performance}
\end{table*}

\begin{table*}[!ht]
	\centering
	\renewcommand{\arraystretch}{1.8}
        \footnotesize
	\caption{Performance comparison using the AAL atlas on ABIDE-I and ADHD-200 datasets.}
	\begin{tabular}{
			m{5.2cm}<{\centering}
			m{1.15cm}<{\centering}
			m{1.15cm}<{\centering}
			m{1.15cm}<{\centering}
			m{1.15cm}<{\centering}
			m{1.15cm}<{\centering}
			m{1.15cm}<{\centering}
			m{1.15cm}<{\centering}
			m{1.3cm}<{\centering}
		}
		\Xhline{1.2pt}
		& \multicolumn{4}{c}{\textbf{ABIDE-I}} & \multicolumn{4}{c}{\textbf{ADHD-200}} \\[1pt]
		\cline{2-9}
		 \multirow{-2}{*}{\textbf{Model}} & \textbf{Acc} & \textbf{Prec} & \textbf{Rec} & \textbf{AUC} & \textbf{Acc} & \textbf{Prec} & \textbf{Rec} & \textbf{AUC} \\[1pt]
		\hline
		SVM \citep{abraham2017deriving}
        & 63.74±3.99                        & 52.35±7.18               & 63.60±6.35              & 69.90±4.53              & 57.96±2.60              & 56.96±5.16               & 57.82±5.16              & 59.40±4.52 \\
	LSTM \citep{dvornek2017lstm}
        & 64.77±4.12                        & 62.12±9.73               & 63.17±7.24              & 68.84±3.77              & 62.10±3.84              & 60.69±4.17               & 58.66±6.06              & 64.24±5.40 \\
		BrainNetCNN \citep{kawahara2017brainnetcnn}
        & 65.31±4.09                        & 60.90±8.56               & 62.88±4.40              & 70.68±5.51              & 61.77±3.78              & 60.11±4.52               & 60.63±7.91              & 60.77±4.01 \\
		\hline
		SwinT \citep{liu2021swin} & 65.69±4.45                        & 58.92±7.45               & 60.53±0.53              & 72.64±4.07              & 66.50±4.60              & 62.64±4.23               & 65.22±9.22              & 66.29±4.45 \\
		BolT \citep{bedel2023bolt}& 69.41±2.15                        & 68.52±4.07               & 66.49±4.22              & 73.30±3.51              & 67.66±3.46              & 68.15±4.05               & 68.36±5.01              & 69.83±3.74 \\
		Com-BrainTF \citep{bannadabhavi2023community} & 70.56±4.42 & 68.53±4.41               & 77.21±4.25              & 75.37±2.96              & 66.87±2.60              & 69.74±4.10               & 70.73±3.59              & 72.06±3.51 \\
		\hline
		MAHGCN \citep{liu2023mahgcn}& 71.31±3.52                        & 69.40±3.27               & 70.09±3.93              & 71.11±2.97              & 69.69±2.49              & 71.36±3.28               & 70.19±4.23              & 71.86±3.31 \\
		
		RGTNet \citep{wang2024rgtnet} & 73.19±3.68                        & 71.45±3.29               & 73.22±3.86              & 72.32±2.68              & 70.06±2.96              & 70.02±3.65               & 73.41±3.70              & 74.75±2.74  \\
		
		PLSNet \citep{wang2023plsnet} & 72.97±3.48                        & 70.87±3.54               & 76.62±3.54              & 75.46±3.25              & 70.53±3.14              & 71.81±2.03               & 72.97±4.86              & 75.45±2.78 \\
		\hline
		STARFormer & \textbf{75.19±2.93}               & \textbf{73.96±2.13}      & \textbf{79.27±3.42}     & \textbf{75.91±2.85}     & \textbf{72.92±2.40}     & \textbf{72.59±2.02}      & \textbf{73.12±3.23}     & \textbf{76.39±2.45} \\
		\Xhline{1.2pt}
	\end{tabular}
	\label{performanceaal}
\end{table*}

The experiments were conducted by using PyTorch based on an NVIDIA RTX 2080Ti GPU. We evaluated the performance of the proposed model using a 10-fold cross-validation, and divided the data into a non-overlapping training set (80\%), a validation set (10\%), and a test set (10\%). All random sampling in our study was performed using simple random sampling without any stratification. The model was trained using the Adam optimizer, with cross-entropy loss as the loss function. To comprehensively evaluate the performance of the model, we employed four commonly used metrics, including accuracy (Acc), precision (Prec), recall (Rec), and area under curve (AUC).

\subsection{Competing Methods}

This study selects three types of baseline models for comparison with STARFormer, namely advanced traditional baselines, transformer-based models, and graph neural networks. Transformer-based models excel at analyzing time series data, while graph neural networks focus on capturing spatial information and modeling spatial structures. All these types of models have been applied in fMRI research, demonstrating strong performance and good generalizability.The architectures, loss functions, and learning rate schedulers for each competing method were adopted from their original papers and subsequently fine-tuned in the experimentation to ensure optimal and competitive performance.

\begin{figure*}[!h]
	\centering
	\includegraphics[width = 1\textwidth]{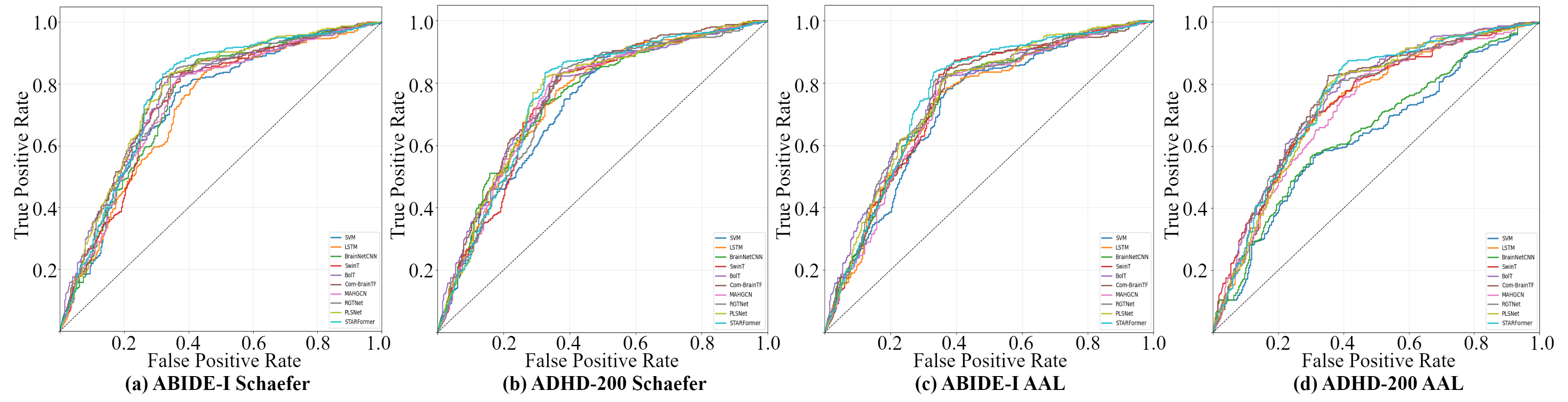}
	\caption{ROC curves for comparing the performance metrics of different methods on ABIDE-I and ADHD-200 datasets using Schaefer and AAL atlases.}
	\label{ROC2}
\end{figure*}

\section{Results}
\subsection{Comparative Studies}
\begin{figure}[h]
	\centering
	\includegraphics[width = 0.45\textwidth]{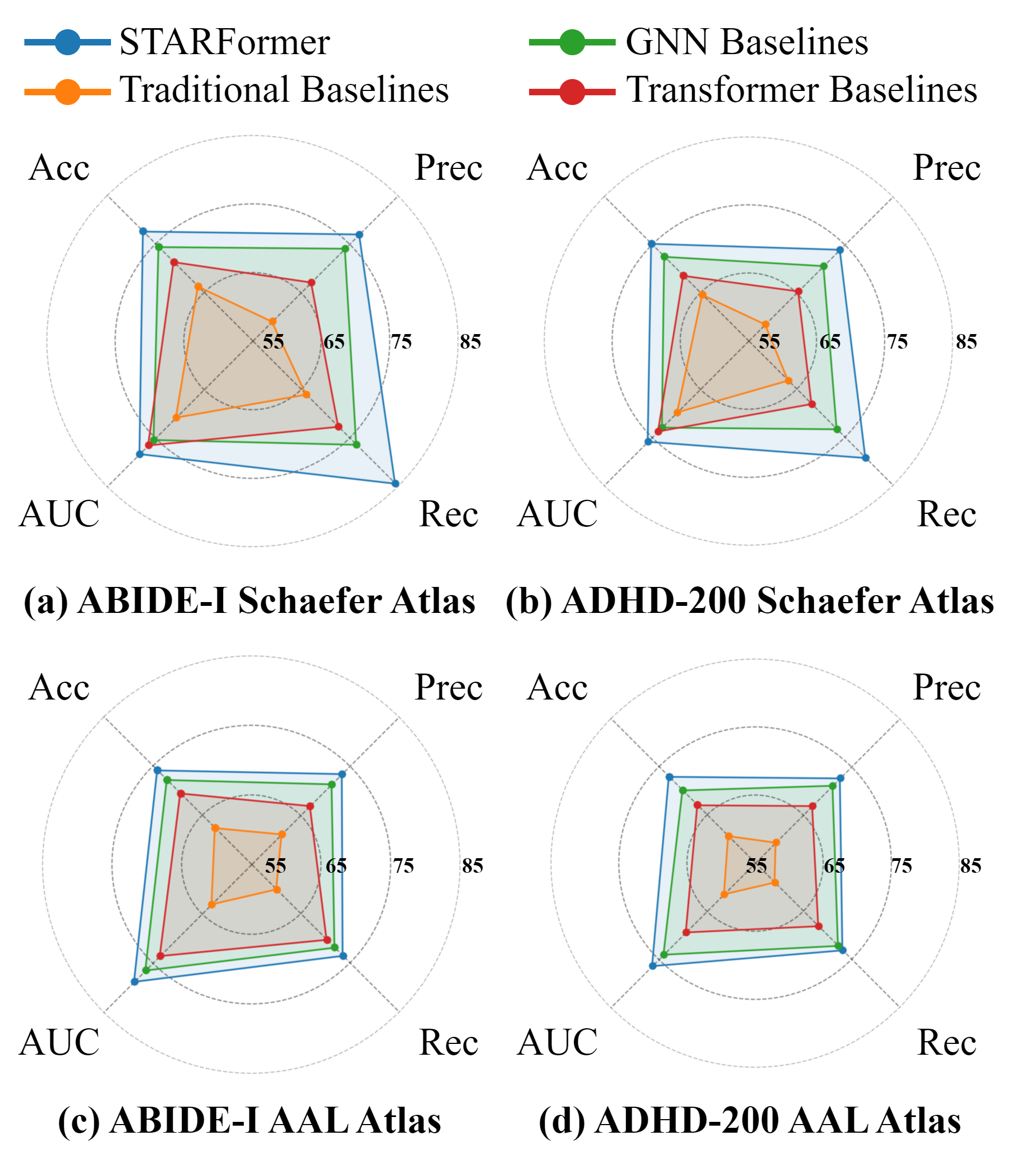}
	\caption{Radar plots for comparing the performance metrics of different methods on ABIDE-I and ADHD-200 datasets using Schaefer and AAL atlases.}
	\label{radar}
\end{figure}

We demonstrate the results of STARFormer in brain disorder diagnosis tasks on ABIDE-I and ADHD-200 datasets using different brain atlases. Table \ref{performance} and Table \ref{performanceaal} present the results of using the Schaefer atlas and the AAL atlas, respectively. It is evident from these tables that our proposed model achieves optimal performance in each metric (p $\leq$ 0.05, Wilcoxon signed-rank test), except for PLSNet offering higher precision on ADHD-200. As expected, different atlases lead to variations in metrics. Specifically, the results using the Schaefer atlas consistently outperform those using the AAL atlas. This is reasonable, as the brain graph based on the former has almost four times the number of ROIs compared to the latter, providing a more comprehensive and detailed information of the brain. Fig.\ref{radar} intuitively shows the comparative performance between different model categories. Each point in the radar plots represents the averaged performance of the methods within the same category, providing an intuitive visualization of the relative strengths of different approaches. The plots clearly demonstrate that STARFormer consistently achieves superior performance across all metrics compared to other baseline methods.

The ROC curve in Fig.\ref{ROC2} shows the trade-off relationship between the true positive rate and the false positive rate of different methods. The AUC value further quantifies the overall discriminative ability of the classifier. The results show that STARFormer exhibits excellent ROC performance under all dataset configurations, indicating that this method has outstanding classification discrimination ability and clinical application potential.

Note that GNN-based methods generally perform better than most models, probably because GNNs have a natural advantage in learning topological information from the brain. Modelling brain functional network as a graph, GNNs can effectively capture connections between brain regions. However, GNN-based methods mainly take static FC as input, which may struggle to capture dynamic information that changes over time. Furthermore, the relative disadvantage of conventional transformer models in fMRI tasks compared to GNN models includes their lack of focus on spatial information. This is because conventional transformers are mainly based on attention mechanisms to capture temporal relationships. In contrast, STARFormer considers the spatial relationships between brain regions while capturing the temporal relationships in the time series. This gives STARFormer a significant performance advantage in brain disorder diagnosis tasks compared to both the transformer baseline models and the GNN baseline models.

\begin{figure*}[!h]
	\centering
	\includegraphics[width = 0.9\textwidth]{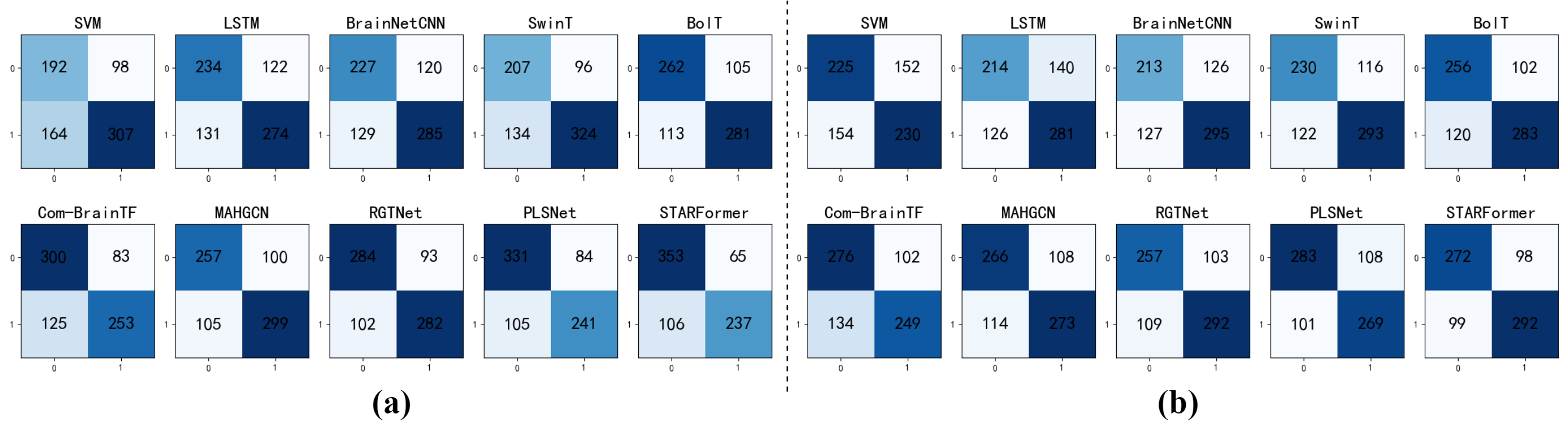}
	\caption{Confusion matrices for all classification methods on ABIDE-1 (a) and ADHD-200 (b) datasets using Schaefer atlases.}
	\label{con1}
\end{figure*}

\begin{figure*}[!h]
	\centering
	\includegraphics[width = 0.9\textwidth]{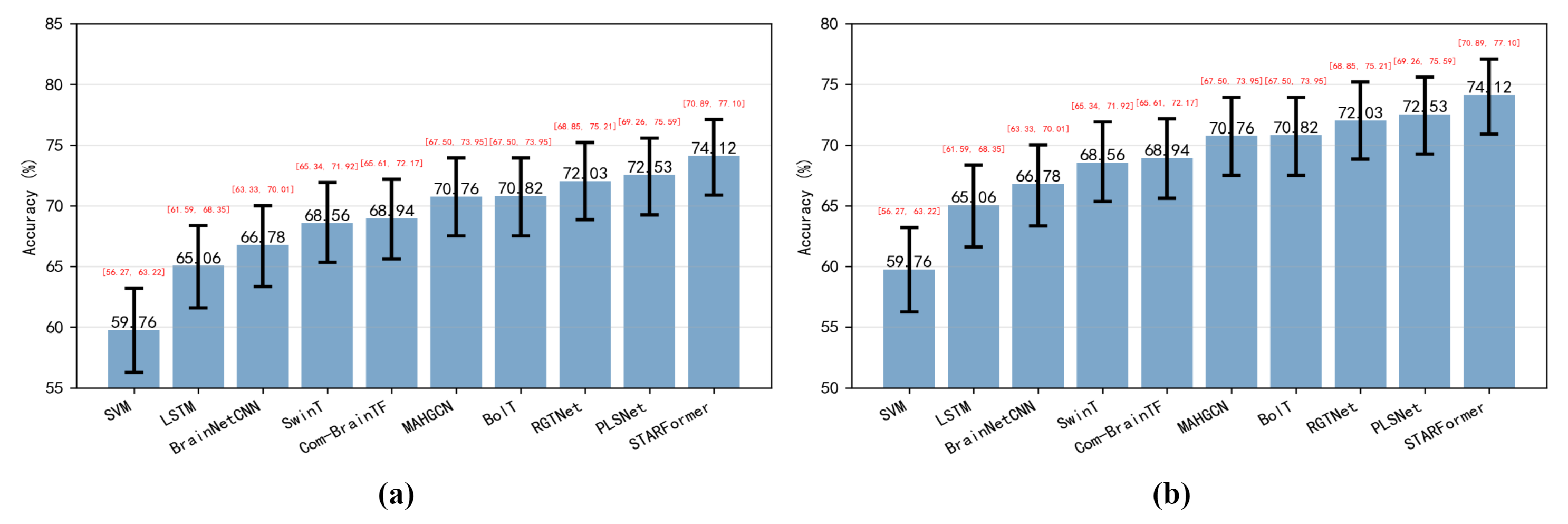}
	\caption{Classification accuracy with 95\% confidence intervals for different methods on ABIDE-1 (a) and ADHD-200 (b) datasets using Schaefer atlases.}
	\label{con2}
\end{figure*}

\begin{figure*}[!h]
	\centering
	\includegraphics[width = 0.9\textwidth]{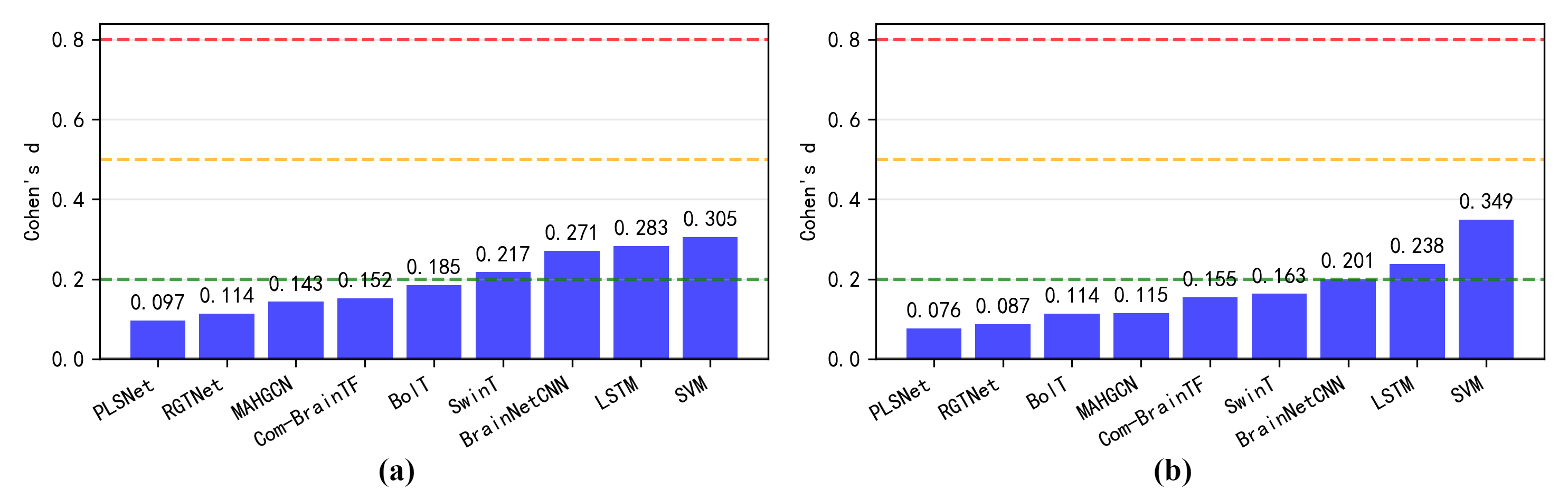}
	\caption{Cohen's d effect sizes comparing different classification methods with STARFormer on ABIDE-1 (a) and ADHD-200 (b) datasets using Schaefer atlases.}
	\label{con3}
\end{figure*}

Fig.\ref{con1} summarizes the classification outcomes for all methods on both datasets via confusion matrices. The stability and reliability of STARFormer are further demonstrated in Fig.\ref{con2}, where the 95\% confidence intervals for accuracy on both ABIDE-1 and ADHD-200 datasets are notably narrow, indicating consistent performance across different cohorts. Regarding effect size analysis, Fig.\ref{con3} reveals that STARFormer achieves substantial improvements over traditional baseline methods, with effect size differences exceeding 0.2, which indicates significant classification advantages. Meanwhile, the performance gap between Transformer-based models and graph neural network methods shows more moderate differences, with effect sizes typically around 0.1, suggesting that both advanced approaches demonstrate comparable capabilities while STARFormer maintains a competitive edge.

\begin{table*}[!ht]
	\centering
	\renewcommand{\arraystretch}{1.8}
        \footnotesize
	\caption{Ablation study of different components on ABIDE-I and ADHD-200 datasets.}
	\begin{tabular}{
			>{\centering\arraybackslash}m{1cm}
			>{\centering\arraybackslash}m{1.2cm}
			>{\centering\arraybackslash}m{1cm}
			>{\centering\arraybackslash}m{1.1cm}
			m{1.1cm}<{\centering}
			m{1.1cm}<{\centering}
			m{1.1cm}<{\centering}
			m{1.1cm}<{\centering}
			m{1.1cm}<{\centering}
			m{1.1cm}<{\centering}
			m{1.1cm}<{\centering}
			m{1.3cm}<{\centering}
		}
		\Xhline{1.2pt}
		\multirow{2}{1.1cm}{\centering\textbf{Variable Window}} & \multirow{2}{1.2cm}{\centering\textbf{Cross-Window Attention}} & \multirow{2}{1.1cm}{\centering\textbf{Spatial Feature}} & \multirow{2}{1.1cm}{\centering\textbf{ROI Analysis}} & \multicolumn{4}{c}{\textbf{ABIDE-I}} & \multicolumn{4}{c}{\textbf{ADHD-200}} \\[2pt]
		\cline{5-12}
		& & & & \textbf{Acc} & \textbf{Prec} & \textbf{Rec} & \textbf{AUC} & \textbf{Acc} & \textbf{Prec} & \textbf{Rec} & \textbf{AUC} \\[2pt]
		\hline
		× & × & × & × & 66.82±2.91 & 66.82±2.16 & 70.81±4.95 & 71.67±6.26 & 65.32±4.31 & 63.11±4.92 & 60.43±6.10 & 65.96±5.51 \\[2pt]
		\hline
		$\surd$ & × & × & × & 71.18±2.99 & 67.62±3.02 & 75.98±5.35 & 74.28±3.58 & 68.06±3.26 & 68.51±4.07 & 68.63±4.37 & 70.83±3.83 \\[2pt]
		\hline
		$\surd$ & $\surd$ & × & × & 73.36±5.03 & 70.39±4.57 & 81.84±5.76 & 75.74±3.58 & 68.90±2.65 & 68.63±3.51 & 70.52±3.53 & 71.54±3.49 \\[2pt]
		\hline
		$\surd$ & × & $\surd$ & × & 74.26±3.93 & 72.11±4.77 & 82.33±5.35 & 76.48±4.57 & 70.79±2.30 & 70.16±3.11 & 71.03±3.97 & 73.24±3.01 \\[2pt]
		\hline
		$\surd$ & $\surd$ & $\surd$ & × & 75.38±4.36 & 73.78±5.18 & 83.13±5.16 & 76.44±4.92 & 71.57±2.94 & 71.64±3.82 & 71.36±4.04 & 74.96±3.22 \\[2pt]
		\hline
		$\surd$ & $\surd$ & $\surd$ & $\surd$(R) & 59.18±3.21 & 63.87±6.67 & 63.55±7.12 & 57.71±3.90 & 58.63±4.32 & 60.33±5.21 & 61.24±5.76 & 59.52±4.00 \\[2pt]
		\hline
		$\surd$ & $\surd$ & $\surd$ & $\surd$(E) & \textbf{77.57±3.70} & \textbf{76.98±3.27} & \textbf{84.38±3.50} & \textbf{78.29±3.23} & \textbf{74.12±2.47} & \textbf{73.37±3.07} & \textbf{73.54±3.27} & \textbf{78.81±2.18} \\[2pt]
		\Xhline{1.2pt}
	\end{tabular}
	\begin{tablenotes}[flushleft] 
		\item[]Notes: The elements of ablation include the variable window, cross-window attention, spatio-temporal feature fusion module (Spatial Feature), and ROI spatial structure analysis module (ROI Analysis). When the ROI spatial structure analysis module is enabled, (R) denotes the random permutation of the ROI sequence, and (E) indicates the ROI arranged based on EC. The results are based on the Schaefer atlas.
	\end{tablenotes}
	\label{ablation}
\end{table*}

\begin{table*}[!ht]
	\centering
	\renewcommand{\arraystretch}{1.8}
        \footnotesize
	\caption{The comparison of performance with different variable window settings on ABIDE-I and ADHD-200 datasets.}
	\begin{tabular}{
			m{3.65cm}<{\centering}
			m{1.35cm}<{\centering}
			m{1.35cm}<{\centering}
			m{1.35cm}<{\centering}
			m{1.35cm}<{\centering}
			m{1.35cm}<{\centering}
			m{1.35cm}<{\centering}
			m{1.35cm}<{\centering}
			m{1.45cm}<{\centering}
		}
		\Xhline{1.2pt}
		\multirow{2}{3.2cm}{\centering\textbf{Variable \\ Window}} 
		& \multicolumn{4}{c}{\textbf{ABIDE-I}} & \multicolumn{4}{c}{\textbf{ADHD-200}} \\[2pt]
		\cline{2-9}
		& \textbf{Acc} & \textbf{Prec} & \textbf{Rec} & \textbf{AUC} & \textbf{Acc} & \textbf{Prec} & \textbf{Rec} & \textbf{AUC} \\[2pt]
		\hline
		\{8, 4, 4, 8\} & 75.29±4.18 & 75.64±3.94 & 82.27±4.01 & 77.09±3.67 & 72.91±2.84 & 68.87±3.54 & 72.62±3.45 & 76.63±3.52 \\[2pt]
		\hline
		\{16, 8, 8, 16\} & 75.86±4.02 & 75.98±4.40 & 84.13±3.16 & 77.51±3.35 & 73.79±2.75 & 69.04±3.18 & 73.51±3.37 & 78.36±3.68 \\[2pt]
		\hline
		\{8, 4, 2, 2, 4, 8\} & 76.78±4.36 & 75.87±3.18 & \textbf{84.41±3.79} & 77.44±3.92 & 73.82±3.51 & 70.45±3.20 & \textbf{74.22±3.86} & 78.32±2.68 \\[2pt]
		\hline
		\{16, 8, 4, 4, 8, 16\} & \textbf{77.57±3.70} & \textbf{76.98±3.27} & 84.38±3.50 & \textbf{78.29±3.23} & \textbf{74.12±2.47} & \textbf{73.37±3.07} & 73.54±3.27 & \textbf{78.81±2.18} \\[2pt]
		\hline
		\{16, 8, 4, 2, 2, 4, 8, 16\} & 74.77±3.00 & 72.25±3.32 & 83.01±3.53 & 76.96±4.33 & 72.53±3.11 & 68.88±3.27 & 71.57±4.31 & 76.45±2.97 \\[2pt]
		\Xhline{1.2pt}
	\end{tabular}
	\label{window_tokens}
\end{table*}

\subsection{Ablation Studies}
We conducted a series of ablation studies to assess the contributions of the design elements in STARFormer. These design elements include the variable window, cross-window attention, spatio-temporal feature fusion module, and ROI spatial structure analysis module. Starting with a standard transformer variant, we gradually introduced the design elements to create ablation variants. For all ablation variants, the architecture and hyperparameters of the components used were matched to those of STARFormer. The standard variant omits all design elements and retains only the fundamental transformer of the temporal branch with a self-attention mechanism. In order to evaluate the contribution of the variable window, a multi-layer transformer block was introduced, and the time series was divided by variable window. Cross-window attention was introduced to form a new variant to assess its contribution. It is important to note that cross-window attention relies on the variable window, thus self-attention was used when the variable window was absent. To evaluate the contribution of spatial features, the spatio-temporal feature fusion module was incorporated into the variants, forming two variants based on the variable window with and without cross-window attention. Finally, the contribution of spatial features together with ROI analysis was evaluated by introducing ROI spatial structure analysis module.

Table \ref{ablation} lists the performance metrics of all ablation variants, showing that the STARFormer model, which incorporates all design elements, achieves the highest performance among all variants. First, the inclusion of the variable window significantly enhances the performance of the transformer, demonstrating that extracting local features at different scales is more practical than directly extracting global representations. Second, we find that using cross-window attention to facilitate information interaction across windows improves performance, indicating the importance of this cross-window attention mechanism for integrating contextual representations of local features across windows. Third, when both temporal and spatial features are extracted and analyzed simultaneously, all performance metrics show improvement. This is because attention to spatial information provides the model with more comprehensive information, thereby having better classification results. Additionally, randomly shuffling the ROI spatial ordering of brain regions leads to a significant drop in model performance due to the disorder among brain regions. Finally, we observe that ranking ROIs within the functional brain network based on EC significantly contributes to improving model performance. We speculate that this is because effective connectivity captures the directional flow of information between different brain regions, and EC distinguishes the ability of nodes to receive and transmit information within the network, making it easier for the model to identify potential features when extracting spatial information. In summary, the use of different ROI spatial ordering for brain regions affects the final results, highlighting the critical importance of spatial information for fMRI data.

\begin{figure}[!h]
	\centering
	\includegraphics[width = 0.5\textwidth]{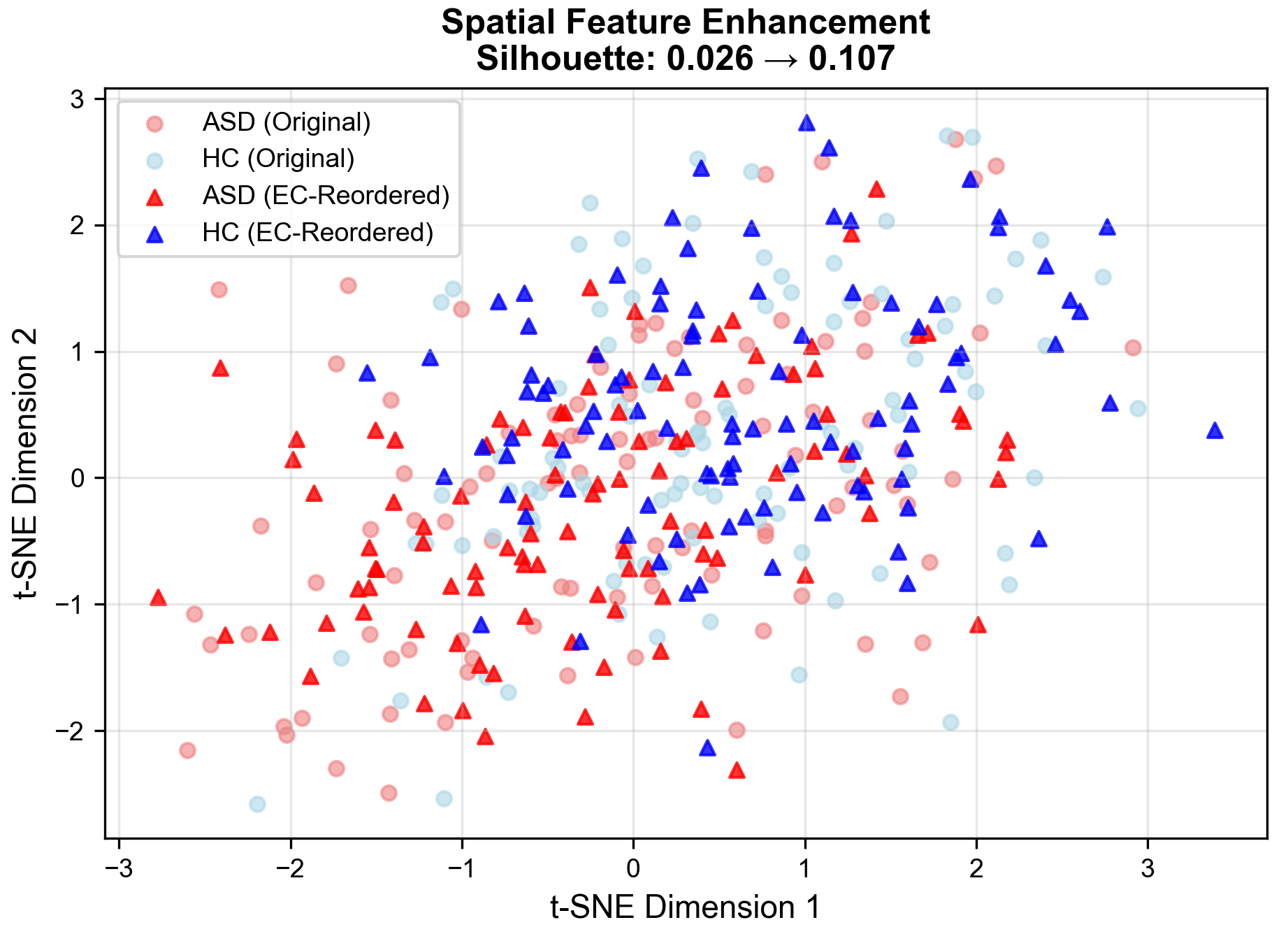}
	\caption{ROI Spatial Structure Analysis Module enhances spatial feature separability through EC reorganization.}
	\label{roi1}
\end{figure}

\begin{figure}[!h]
	\centering
	\includegraphics[width = 0.7\textwidth]{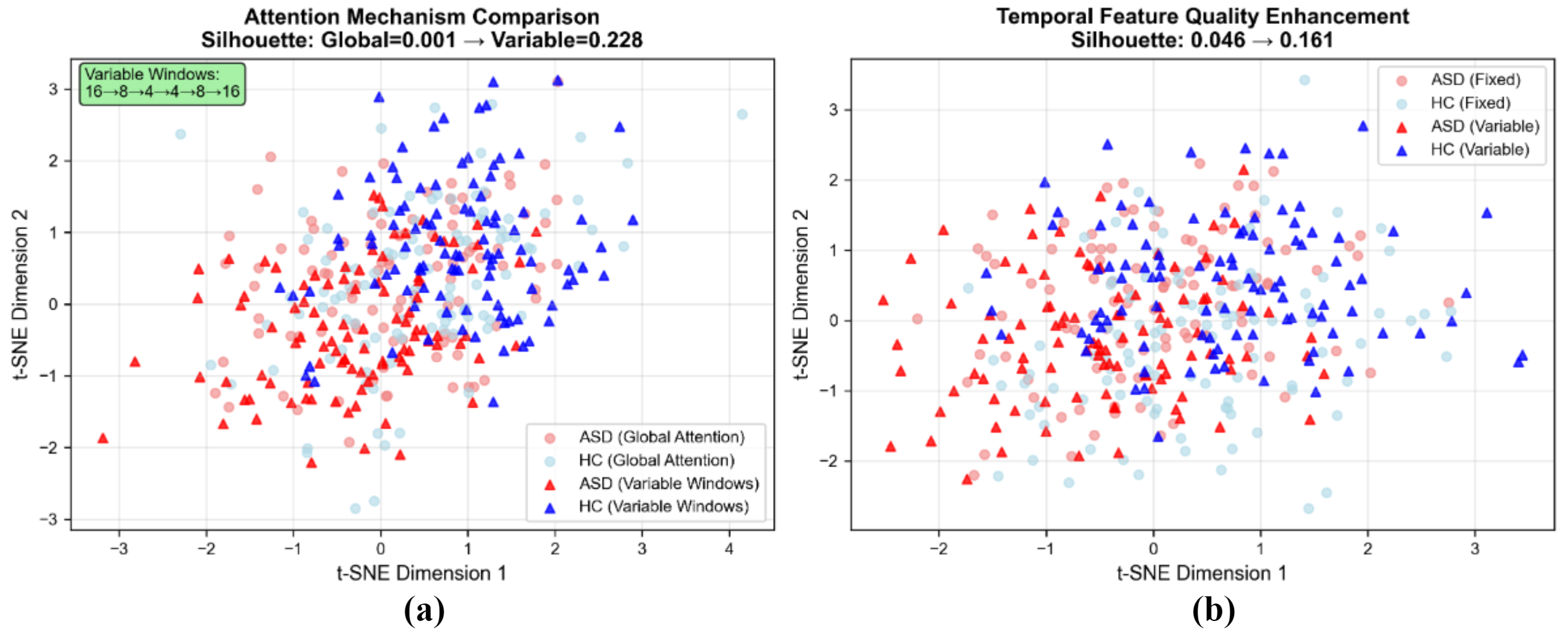}
	\caption{Temporal Feature Reorganization Module. (a) Variable window attention vs. global attention comparison. (b) Temporal feature enhancement via t-SNE visualization.}
	\label{roi2}
\end{figure}

\begin{figure}[!h]
	\centering
	\includegraphics[width = 0.7\textwidth]{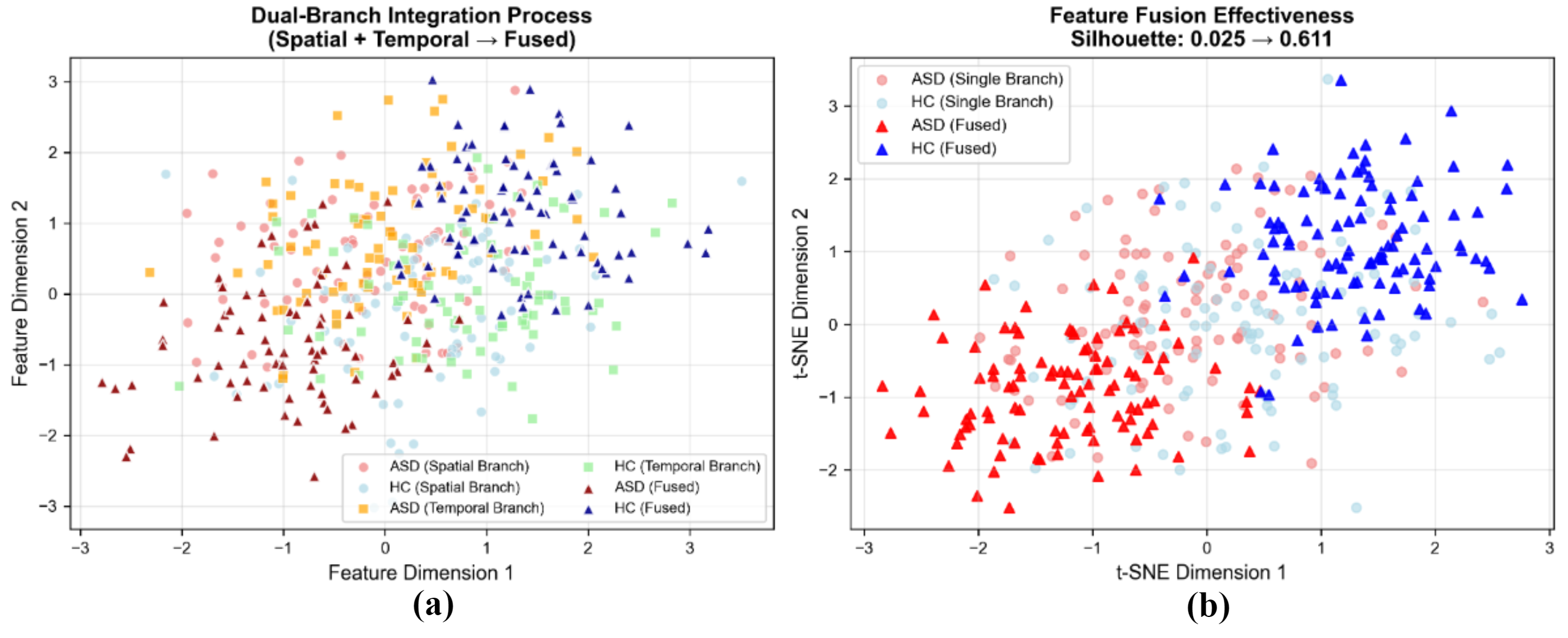}
	\caption{Spatio-Temporal Feature Fusion Module. (a) Dual-branch integration process from spatial and temporal features to fused representation. (b) Feature fusion effectiveness.}
	\label{roi3}
\end{figure}

We implement t-SNE-based feature visualization with silhouette coefficient quantification for each module. Fig.\ref{roi1} demonstrates how the ROI Spatial Structure Analysis Module enhances spatial feature separability through EC reorganization. Fig.\ref{roi2} shows how the Temporal Feature Reorganization Module improves temporal pattern discrimination via variable window mechanisms. Fig.\ref{roi3} illustrates how the Spatio-Temporal Feature Fusion Module creates superior integrated representations by combining dual-branch features.

\subsection{Window Setting}
We evaluated the STARFormer variants obtained from different windows applied to fMRI time series. The Table \ref{window_tokens} lists the performance metrics of different variants of STARFormer variable windows using the ABIDE-I dataset and the ADHD-200 dataset based on the Schaefer atlas. We observed that the performance advantage of the variable window maximizes when the number of window tokens is \{16, 8, 4, 4, 8, 16\}. This may be because too few layers lead to insufficient learning capacity of the model, preventing it from extracting enough features, while too many layers may cause the model to overfit, resulting in good performance on the training set but poor generalization to the test set. Additionally, since the number of window tokens determines the size of the window, smaller window tokens allow the model to focus more on local features, enabling it to capture detailed information better. However, smaller window tokens will limit the ability of the model to perceive global information, making it difficult to understand broader feature dependencies, especially in tasks that require capturing cross-regional or long-term dependencies. In contrast, larger window tokens are capable of capturing global features.

\begin{figure}[!h]
	\centering
	\includegraphics[width = 0.5\textwidth]{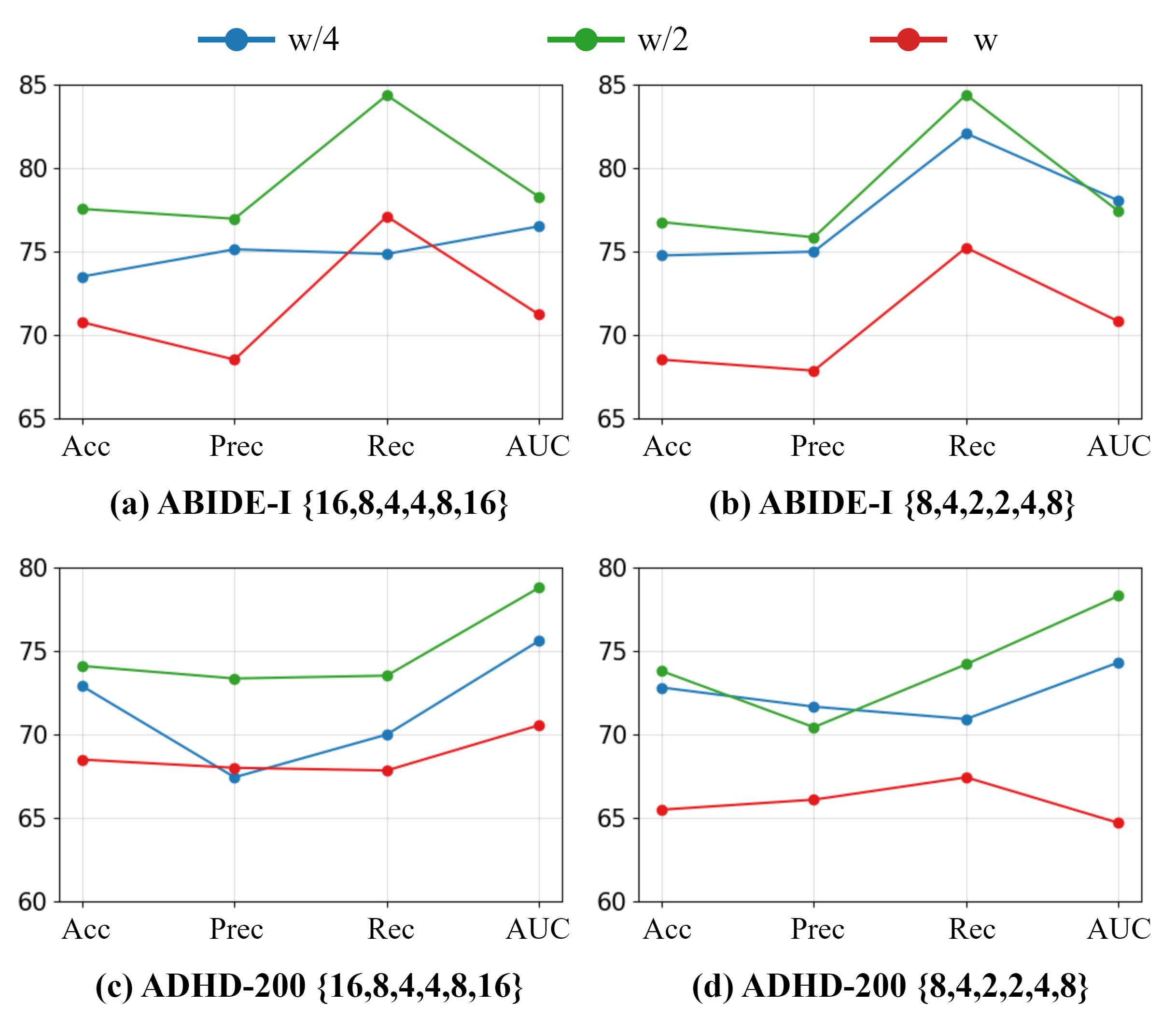}
	\caption{The comparison of the performance of STARFormer variants with the configurations of different extended window based on ABIDE-I and ADHD-200 datasets.}
	\label{w}
\end{figure}

As shown in Fig.\ref{w}, we examined the impact of the size of the extended window under a specific setting of variable window on model performance, training with extended window of sizes $w/4$, $w/2$, and $w$. We found that performance exhibits moderate variation with changes in the extended window token size. In other words, the model tends to achieve optimal performance with a window size of $w/2$. Overall, the size of the window needs to be set with a balance between the needs for local and global information. We found that both datasets typically achieved optimal or near-optimal performance when the number of window tokens for the variable window was \{16, 8, 4, 4, 8, 16\} and the extended window size was $w/2$. This result indicates a degree of reliability in the introduction of window-related design elements in STARFormer. 

\begin{figure*}[!h]
	\centering
	\includegraphics[width = 1\textwidth]{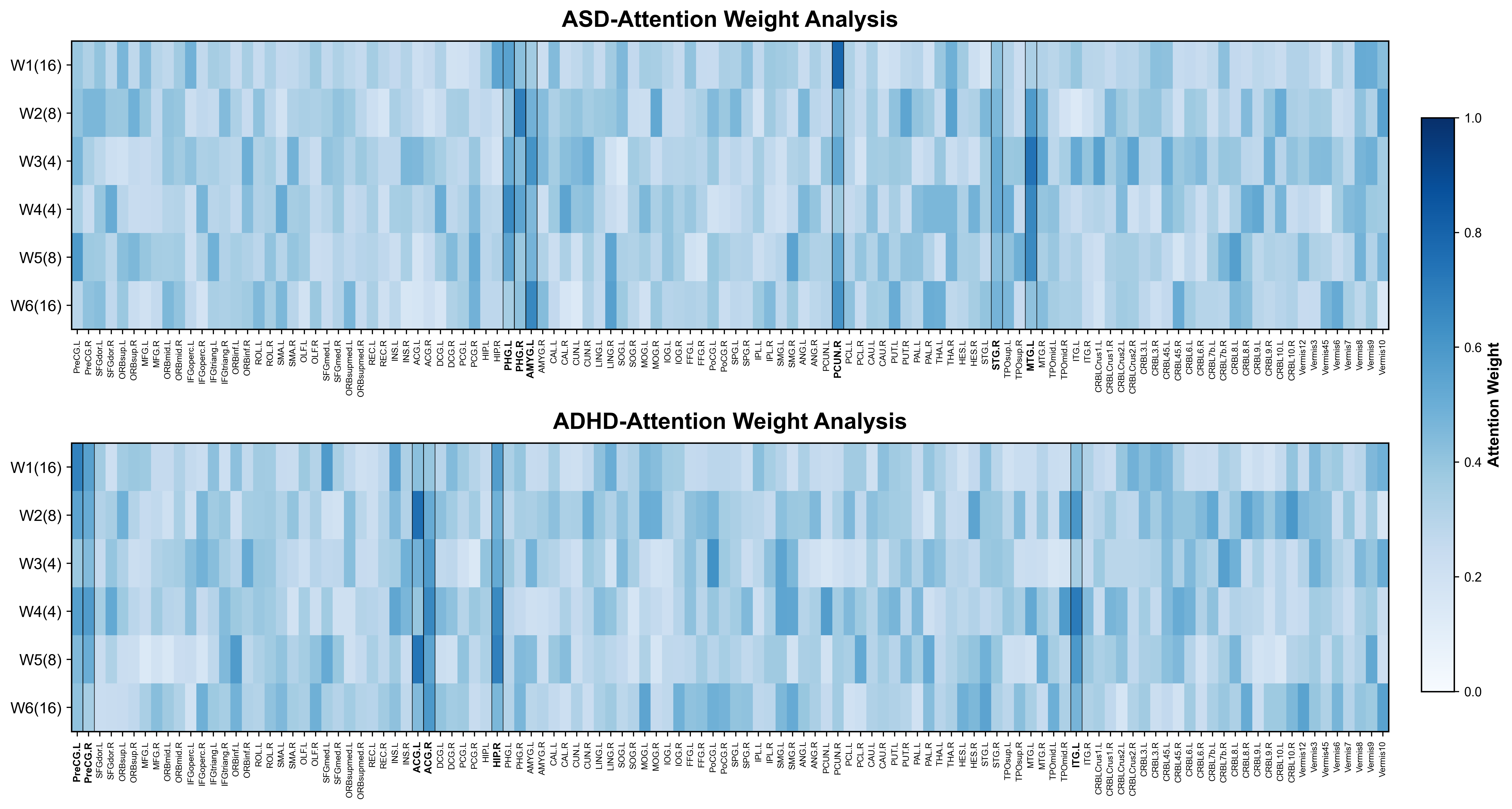}
	\caption{Temporal attention analysis showing ROI importance across temporal branch.}
	\label{p1}
\end{figure*}

\begin{figure*}[!h]
	\centering
	\includegraphics[width = 1\textwidth]{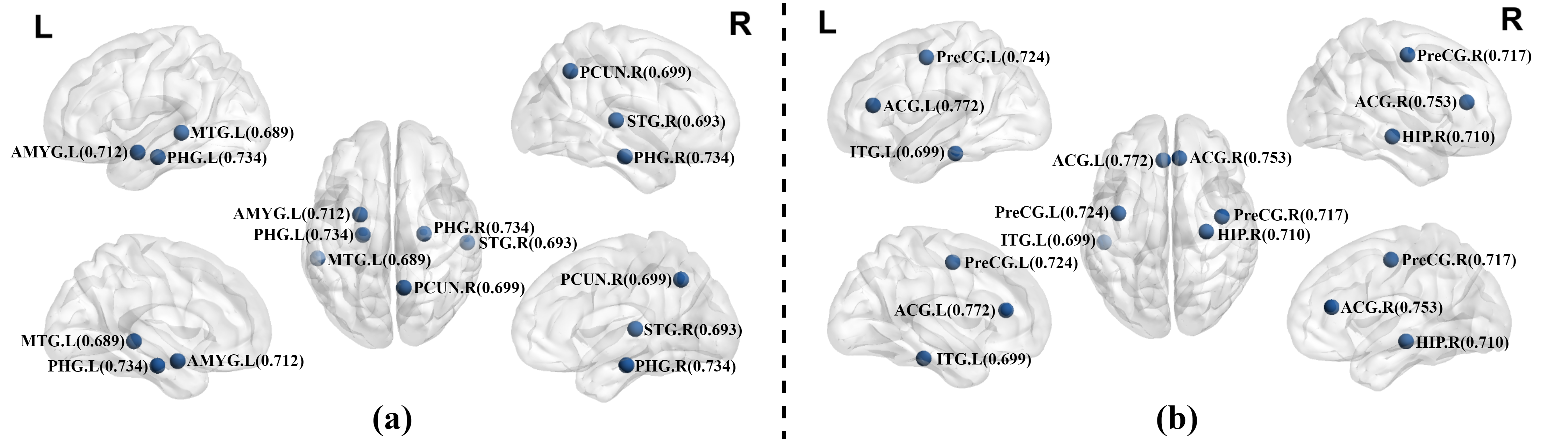}
	\caption{Top 5\% ROIs which are most important for ASD classification (a) and ADHD classification (b) according to importance scores based on the attention matrix in the transformer block of STARFormer.}
	\label{brain}
\end{figure*}

\section{Discussion}
\subsection{Interpretability Analysis}

We used interpretability techniques to further analyze brain regions important for ASD and ADHD in the STARFormer model. In the temporal branch, the attention matrices (i.e., the Attention(Q, K, V) defined in Equation~(\ref{eq:attention})) of each layer are averaged. In the spatial branch, the attention matrices are multiplied by the feature activation values. Subsequently, the results of the temporal and spatial branches are each aggregated by rows and normalized to obtain attention importance scores for temporal and spatial dimensions. Finally, by combining these two scores through weighted summation, a comprehensive importance score is obtained to identify the most influential ROIs.

We implement a systematic ROI identification procedure using multi-head attention weight analysis. Fig.\ref{p1} shows the temporal attention analysis, where we extract attention weights from STARFormer's temporal branch architecture across different temporal windows and brain regions organized by functional networks. This heatmap visualization reveals which ROIs receive the highest attention weights during classification, providing a systematic ranking of ROI importance.

As shown in Fig.\ref{brain}, we present the top 5\% most influential ROIs for diagnosing ASD and ADHD. A manual review confirmed that all identified ROIs are consistent with the previous literature, linking them with the neural manifestations of ASD and ADHD. For instance, functional abnormalities in the parahippocampal gyrus may impair the ability of individuals with ASD to adapt or recall relevant information in social settings, affecting their social behavior \citep{monk2009abnormalities}. Siminarly, functional abnormalities in the amygdala can lead to anxiety or avoidance in social situations for those with ASD \citep{kleinhans2009reduced}. In addition, functional abnormalities in the precentral gyrus have been associated with hyperactivity symptoms in individuals with ADHD \citep{lei2015functional}, while abnormalities in the inferior temporal gyrus can result in attention deficits or distractibility during complex visual tasks \citep{kobel2010structural}. Overall, this demonstrates that the STARFormer effectively captures brain activation patterns in both healthy individuals and those with brain disorders.

\subsection{Limitation and Future Work}
While our proposed STARFormer shows significant improvement over existing computer-aided diagnosis methods for brain disorders, several issues should be considered in future work. First, we used node centrality measures to extract spatial information for ROIs rather than directly using the topological information of the brain network. Future research could explore using graph encoding to model brain topology. Second, auxiliary information about patients, such as personal details or scanning protocols, was not included as input to the model. Recent studies suggest that these phenotypic data complement imaging data and may improve the diagnosis of brain disorders \citep{Zhang2024Preserving}. Therefore, it is reasonable to expect that integrating phenotypic information could further improve the classification performance of STARFormer. Lastly, considering the challenges in data acquisition in clinical settings, many subjects will have partially labeled fMRI data. Thus, strategies using semi-supervised or unsupervised learning could be considered for training.

\section{Conclusion}
In this study, we introduce STARFormer, an advanced transformer-based framework for diagnosing brain disorders that effectively integrates spatial structure analysis with temporal feature learning. Through its novel architecture, STARFormer successfully addresses key limitations of existing methods by simultaneously capturing the intricate spatial relationships between brain regions and both local and global temporal patterns in fMRI data.

Comprehensive empirical evaluations of the ABIDE-I and ADHD-200 datasets demonstrate that STARFormer achieves superior performance in both ASD and ADHD classification tasks, significantly outperforming existing state-of-the-art methods. The framework's ability to identify specific ROIs related to brain disorders aligns with established neurological findings, validating its potential for clinical applications. These results show that STARFormer advances the technical frontier of brain disorder diagnosis. Future research may explore the adaptability of the framework to other neurological disorders and its potential integration into clinical decision support systems.

The modular design and efficient architecture of STARFormer make it suitable for deployment in real-time or large-scale clinical settings. With integrated automated preprocessing and inference, the model can accelerate processing to support rapid screening and risk assessment. Future work can focus on model compression, lightweight deployment, and adaptation to heterogeneous data to facilitate its practical application in hospital information systems or neuroimaging cloud platforms.

\section*{CRediT authorship contribution statement}
\textbf{Wenhao Dong:} Conceptualization, Methodology, Software, Writing – original draft, Writing – review \& editing. \textbf{Yueyang Li:} Conceptualization, Methodology, Software, Writing – original draft, Writing – review \& editing. \textbf{Weiming Zeng:} Conceptualization, Supervision. \textbf{Lei Chen:} Software, Formal analysis. \textbf{Hongjie Yan:} Validation, Writing – review \& editing. \textbf{Wai Ting Siok:} Validation, Writing – review \& editing. \textbf{Nizhuan Wang:} Conceptualization, Writing – review \& editing, Supervision, Project administration

\section*{Declaration of competing interest}
The authors declare that they have no known competing finan
cial interests or personal relationships that could have appeared to
 influence the work reported in this paper.

\section*{Acknowledgments}
This work was supported by the National Natural Science Foundation of China (grant number 31870979), The Hong Kong Polytechnic University Faculty Reserve Fund (Project ID: P0053738), The Hong Kong Polytechnic University Start-up Fund (Project ID: P0053210), an internal grant from The Hong Kong Polytechnic University (Project ID: P0048377), The Hong Kong Polytechnic University Departmental Collaborative Research Fund (Project ID: P0056428) and The Hong Kong Polytechnic University Collaborative Research with World-leading Research Groups Fund (Project ID: P0058097).

\section*{Data availability}
The data that support the findings of this study are publicly available from ABIDE-I (\url{https://fcon\_1000.projects.nitrc.org/indi/abide/abide\_I.html}) and ADHD-200 (\url{http://preprocessed-connectomes-project.org/adhd200}).

\bibliographystyle{unsrt}  
\bibliography{references}

\end{document}